\documentclass[10pt]{article}
\usepackage[cp866]{inputenc}
\usepackage[russian]{babel}

\begin{document}


\newcommand{\N}{N\raise.7ex\hbox{\underline{$\circ $}}$\;$}

\begin{center}
{\bf

\thispagestyle{empty}

BELARUS NATIONAL ACADEMY OF SCIENCES

B.I. STEPANOV's INSTITUTE OF PHYSICS

}
\end{center}

\vspace{50mm}

\begin{center}

{\bf Red'kov V.M.\footnote{E-mail: redkov@dragon.bas-net.by}}

\vspace{10mm}

{\bf ON  WKB-QUANTIZATION FOR KEPLER PROBLEM \\ IN   EUCLIDE, RIEMANN AND  LOBACHEVSKY 3-SPACE}

\end{center}

\begin{quotation}

Quantum mechanical WKB-method is elaborated for the known quantum  Kepler problem
in curved 3-space models Euclide, Riemann  and Lobachevsky in the framework of the
 complex variable function theory. Generalized Schr\"{o}din\-ger, Klein-Fock
 hydrogen atoms are considered. Exact energy levels are found and their exactness
 is proved on the base of exploration  into   $n$-degree terms of the WKB-series.
Dirac equation is solved too,  but only approximate energy spectrum is established.

\vspace{10mm}

18 page, 68 references

\end{quotation}

\newpage

\subsection*{1. Introduction}

It is well known that energy spectrum of the hydrogen atom had been calculated  long
before  creating  the consistent   quantum mechanical theory.
It was established that the  Bohr-Sommerfeld rules, basis of the "old"\hspace{2mm}
quantum mechanics even without rigorous mathematical foundation, are closely referred to
the so-called WKB-approximation it the consistent quantum theory [1-45].

Looking for solvable  models in  the framework of
"new" \hspace{2mm} quantum theory, some coolness towards  approximate
 methods and any achievements of the Bohr-Sommerfeld mechanics  was inevitable.
But now and then the same  question arises in the literature: why in the case of hydrogen
atom  the Bohr-Sommerfeld rule leads to the known exact energy spectrum.
Also,  from time to time in the  literature one can face  statement of the sort:
in a potential $ \varphi $ the Bohr-Sommerfeld quantization gives an exact result
$\epsilon_{n} (\varphi)$.

In the present work we turn again to a hydrogen atom   but now placed concurrently in three
different curved space backgrounds: Euclide $E_{3}$ (zero curvature), Riemann $S_{3}$ (positive
constant curvature),  and Lobachevsky $H_{3}$ (negative constant  curvature).
We have considered models based on Schr\"{o}dinger, Klein-Fock equations in these three space models
and shortly analyzed a hydrogen atom described by Dirac equation in $E_{3}$-space.
From the very beginning it should be noted that these three hydrogen atom models prove themselves
 essentially
different from mathematical and physical standpoint [46-68], and their study
manifestly adds to general geometry-physics ground and to development of the quasi-classical method itself
.

For the first time the  hydrogen atom in a curved space-time was considered by
E.  Schr\"{o}dinger [46].  In this  work he elaborated  the so-called factorization
method as applied to  some quantum mechanical  eigenvalues problems, in particular the harmonic
oscillator and the  hydrogen atom.
However, the method fails to function for a continuous part of the hydrogen energy spectrum.
An idea to alter the task in order to obtain only discrete energy levels  had arisen.
Such a full discreteness is achieved by mere placing the  hydrogen atom into a Riemannian
3-space of constant  positive curvature $S_{3}$, because this 3-space is compact and therefore
it might reproduce the effect of a finite dimensions box.

In hyper-spherical  coordinates $(\chi, \theta, \phi )$  of  $S_{3}$-model
$$
dl^{2} = d \chi^{2} + \sin^{2}\chi \; (d\theta^{2} + \sin^{2} \theta d\phi^{2}) \;
\eqno(1.1a)
$$

\noindent
the quantum mechanical Hamiltonian of the system has the form
$$
H = -{1 \over 2} \;
{1 \over \sqrt{g}} \; {\partial \over \partial x^{\alpha}} \; \sqrt{g}
\; g^{\alpha \beta}\; {\partial \over \partial x^{\alpha}} \;
-\; {e \over \tan \chi }  \;  .
\eqno(1.1b)
$$

\noindent
Let us use the following  notation:
the curvature radius $\rho$ is a length unit, $M$ is an electron mass,
$\hbar^{2} /M\rho^{2}$ is an energy unit,
$e = {\alpha \over  \rho } /  { \hbar^{2} \over M \rho^{2} } $ is an effective constant
characterizing intensity of the Coulomb interaction. The sign at $\;\;(e / \tan \chi) $
hand-picked in eq. (1.1b)
As expected  the energy spectrum of the task turns out to be discrete in full, and as shown
 in [46] it is
$$
\epsilon_{n} = - {e^{2} \over 2 n^{2} } + {1 \over 2} \; (n^{2} -1) \; ,
 \;\;  n = 1,2, 3, ...      \;.
\eqno(1.1c)
$$

\noindent  The level $\epsilon_{n} $    displays  $n^{2}$-degeneration  similar to that existing
in ordinary hydrogen atom theory. The number of negative energy levels is finite and
it equals
$$
x_{0}-1, \qquad \mbox{where} \qquad
x_{0}^{2} = {1 \over 2} (1 + \sqrt{1 + 4e^{2} }),
\eqno(1.1d)
$$

\noindent
if  $x_{0}$ is integer; at this there exists a  zero level  $\epsilon = 0$.
When  $x_{0}$ is fractional the number of negative energy levels equals
an entire part of  $x_{0}$. As long as    $x_{0} \geq 1 $ holds,
no matter how small is $e$ (arbitrary small Coulomb  attraction) always there will exists
at least one negative energy level. In other words always there will exist at least one
bound state of the system. It should be especially mentioned that in a
$E_{3}$-based  model of the  hydrogen atom,
no matter how small the  intensity of the Coulomb attraction is
there exists an infinite set of energies being concentrated  near a zero point.

A hydrogen atom model in Lobachevsky space $H_{3}$ turns out to reveal  still more  uncommon
properties [47].
In hyper-spherical coordinates of $H_{3}$-space
$$
dl^{2} = d \chi^{2} + \sinh^{2}\chi (d\theta^{2} + \sin^{2} \theta d\phi^{2})
\eqno(1.2a)
$$

\noindent
the quantum mechanical Hamiltonian of the system  looks
$$
H = -{1 \over 2} \;
{1 \over \sqrt{g}} \; {\partial \over \partial x^{\alpha}} \; \sqrt{g}
\; g^{\alpha \beta}\; {\partial \over \partial x^{\alpha}} \;
- \; {e \over \tanh \chi }  \;  .
\eqno(1.2b)
$$

\noindent In [47] an energy spectrum and corresponding wave functions
$\;\Psi _{\epsilon lm}(\chi,\theta,\phi) $ were  found.
The energy spectrum has a discrete and continuous parts; and the  discrete part
contains a \underline{finite} number of energy levels
$$
- \; {e^{2} \over 2} \leq \epsilon \leq ({1\over 2}  - e ) : \qquad
\epsilon_{n} = - {e^{2} \over 2 n^{2} } - {1 \over 2} \; (n^{2} -1) \; ,
 \;\;  n = 1,2, 3, ...,N \; .
\eqno(1.2c)
$$

\noindent At this the level $\epsilon_{n}$ is $n^{2}$-degenerate in a way similar to
the common hydrogen atom in the flat space.
At  $\epsilon \geq ({1\over2} - e)$ the  energy spectrum is continuous.
The number $N$ of discrete energy levels equals to  $(\sqrt{e} -1)$ at integer $e$,
and $N$ equals to an entire part of  $\sqrt{e}$ if  $\sqrt{e}$ is fractional.
From the above it follows that along with decreasing $e$-parameter the number $N$ gradually
decreases too, and the point  $e=1$ represents a crucial moment after which no bound states in the
system exists. In other words, in the $H_{3}$-space, the hydrogen atom is possible bound system
only at $e>1$.

Here one remark should be added. The distinctive features of the Kepler quantum
problem in curved space backgrounds described above show potential value of these systems (and
similar to them) from another view point. Any task, firstly being formulated as an old
quantum problem in a new space background can be re-formulated  as a certain special quantum
problem in an old flat space background. So, there exists a systematic way based on geometry
to reproduce some  quantum mechanical potentials with  physically interesting properties.
Such a technique  might be understood as  a way to reproduce alternatively  the usual  potential approach
to quantum mechanics through a geometry. Obviously, such an approach was discussed and
used in the  literature (for example, see [66]).

\subsection*{
2. WKB-quantization and hydrogen atom in  $E_{3}$-space}

\begin{quotation}

In  Section 2, general mathematical
structure of $n$-order term  $Q_{n}$ in   WKB-series is established.
It is proved that all terms $Q_{n}, n \geq 2 $ do not contribute into an exact quantum
mechanical quantization  condition.
In the light of this,  success of the  old quantum mechanics is fully rationalized.

\end{quotation}

Let us consider again  a non-relativistic hydrogen atom in Euklide space model $E_{3}$.
Schr\"{o}dinger's equation after separation of variables by a substitution
  $\Psi = R(r) \; Y_{lm}(\theta,\phi)$ leads to
$$
{1 \over r^{2}} {d \over dr } r^{2}  \; {d \over dr } \; R \; + \;
[ \; {2M \over \hbar^{2} } (\epsilon + {\alpha \over r}) \;-\;
{l(l+1)\over r^{2}} \; ] \; R \;= 0 \; .
\eqno(2.1)
$$

\noindent Let $t$ be a new  variable:
$t = \ln  r , \; r = e^{t}, \;  t \in ( -\infty , + \infty )$, then
$$
{d^{2} \over d t^{2}} R \; + \; {d \over d t} R \; + \;
[\; {2 \epsilon M\;  e^{2t} + 2 \alpha M\; e^{t} \over \hbar^{2}}\;  - \;
l(l+1) \;  ] R = 0  \; .
\eqno(2.2a)
$$

\noindent One can exclude the  first derivative term by a substitution $R(t) = e^{-t /2} \; S(t)$
$$
( \; {d^{2} \over d t^{2}}  \; + \;
{\Pi^{2} \over \hbar^{2}}\; ) \; S = 0 \; , \;
$$
$$
\Pi^{2} = 2 \epsilon M\;  e^{2t} + 2 \alpha M\; e^{t} - L^{2} \; , \qquad
L^{2} = \hbar^{2}(l+ {1 \over 2})^{2} \;
\eqno(2.2b)
$$

\noindent and further
$$
S(t) = exp \; [ {i\over \hbar} \; \int \; Q(t) \; dt  \; ] \; ,  \qquad
{\hbar \over i }\; {d \over d t} Q \;  + \; Q^{2 }  \; -  \; \Pi^{2} = 0\; .
\eqno(2.3)
$$

\noindent Now let us expand the function  $Q(t)$ into a series in terms of ($\hbar / i)^{n}$
$$
Q(t) = \sum_{n=0}^{\infty}\;
 [ \; ({\hbar \over i})^{n} Q_{n}(t) \; ] \;.
\eqno(2.4)
$$

\noindent Substituting this formal decomposition into eq. (2.3b), we  will have
$$
(Q_{0}^{2} - \Pi^{2}) \; + \; \sum_{n=1}^{\infty} \;\;
({\hbar \over i})^{n} \;  [ \; {d \over dt}\; Q_{n-1} \; + \;
\sum_{k=0}^{n} \; Q_{n-k}\; Q_{k} \;  ] = 0 \; ,
\eqno(2.5)
$$

\noindent from which it follows
$$
Q_{0}^{2} = \Pi^{2} \; , \;\; Q_{n} = -\; {1 \over 2 Q_{0}} \;
( \; {d \over dt}\; Q_{n-1} \; + \;
\sum_{k=1}^{n-1}  \; Q_{n-k} \; Q_{k} \; ) = 0 \; , \;\;\; n = 1,2,...
\eqno(2.6a)
$$

\noindent Starting from the known $Q_{0}$, the relations  (2.6a) enable us to calculate
generally speaking  all remaining  $Q_{n}$; some first such terms are
(below the symbol $'$ will denote   $(d /dt)$-derivative):
$$
Q_{0}^{2} = \Pi^{2} \qquad \Longrightarrow \qquad   Q_{0} = \pm \sqrt{\Pi^{2}} \; ,
$$
$$
 Q_{1} = - \; {1 \over 2 Q_{0}} \; Q'_{0}  \; , \qquad
Q_{2} = - \; {1 \over 2 Q_{0}}\; (\; Q'_{1} \; + Q_{1}^{2} \; ) \;, \;  ...
\eqno(2.6b)
$$

Now we are at the point to formulate general quantization condition:

\begin{quotation}

We will assume that the wave function  $S(t)$ corresponding to
a bound state, being considered as a function of a complex variable $t$,
has a finite numbers of zeros in the complex plane, which are allocated at real $r$-axis
between classical turning points. According the known theorem in complex
variable functions theory, the number of such zeros of $S(t)$ within certain domain
can be calculated through an integral of the logarithmic derivative $(\ln S(t))'$
along a contour bounding that domain:
\end{quotation}
$$
{1 \over 2\pi i} \; \oint_{{\cal L}} \;
[ \; {d \over d t} \; \ln S(t) \; ] \; dt = n \; ,
\eqno(2.7)
$$

\noindent
Taking into account  (2.3), the quantization condition (2.7)
will take the form
$$
\oint_{{\cal L}}  \; Q(t) \; dt = 2\pi \hbar \;n \; ,
\eqno(2.8)
$$

\noindent
from which substituting a series (2.4) instead of $Q(t)$, we arrive  at
$$
\sum_{n=0}^{\infty} \;  [ \; ({\hbar \over i})^{n} \;
\oint _{\cal L} \; Q_{n}(t) \;dt \;  ] \; = 2 \pi \hbar \; n \; .
\eqno(2.9a)
$$

\noindent
It should be especially emphasized  that  relations (2.7), (2.8) and (2.9a)
are precise mathematical conditions without any approximation.

Below,  at calculating separate contour integrals in (2.9a)
we  will be  needing  to use a complex variable  $z = e^{t}$,
so instead of  (2.9a) one has
$$
\sum_{n=0}^{\infty} \; [ \; ({\hbar \over i})^{n} \;
\oint _{\cal L} \; Q_{n}(z) \;{dz \over z} \;   ] \; =
2 \pi \hbar \; n \; .
\eqno(2.9b)
$$


All terms $z^{-1}\; Q_{n}(z)$ in the WKB-series (2.9b)   are single valued and analyti\-cal  functions
of the variable $z$  just in 3-connected domain of $z$-plane, bounded by three contours
${\cal L}^{all} = {\cal L}_{0} + {\cal L}_{\infty} + {\cal L}$ (see. Fig. 1; dashed line
 designates a special cut needed to allow for existence of two branch points of $ Q_{0} =
 \sqrt{\Pi^{2}}$;  ${\cal L}_{0}$ is any enough small contour around zero point  $z = 0$;
  ${\cal L}_{\infty}$ is any enough large  contour around $\infty$-point):

\vspace{+5mm}
\unitlength=0.6mm
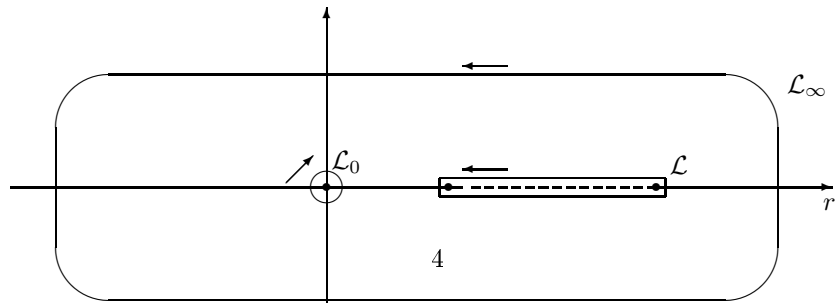
\begin{picture}(160,40)(-120,0)
\special{em:linewidth 0.4pt}
\linethickness{0.4pt}

\put(-90,0){\line(+1,0){98}}
\put(+52,0){\vector(+1,0){40}}

\put(+8,0){\line(+1,0){2}}
\put(+12,0){\line(+1,0){2}}
\put(+15,0){\line(+1,0){2}}
\put(+18,0){\line(+1,0){2}}
\put(+21,0){\line(+1,0){2}}
\put(+24,0){\line(+1,0){2}}
\put(+27,0){\line(+1,0){2}}
\put(+30,0){\line(+1,0){2}}
\put(+33,0){\line(+1,0){2}}
\put(+36,0){\line(+1,0){2}}
\put(+39,0){\line(+1,0){2}}
\put(+42,0){\line(+1,0){2}}
\put(+45,0){\line(+1,0){2}}
\put(+48,0){\line(+1,0){2}}
\put(+51,0){\line(+1,0){2}}
\put(+54,0){\line(+1,0){2}}

\put(+90,-5){$r$}
\put(-20,-40){\vector(0,+1){80}}

\put(+5,-2){\line(+1,0){50}}
\put(+5,-2){\line(0,+1){4}}
\put(+55,+2){\line(-1,0){50}}
\put(+55,-2){\line(0,+1){4}}

\put(+7,0){\circle*{2}}
\put(+53,0){\circle*{2}}
\put(-20,0){\circle*{2}}
\put(-20,0){\circle{7}}

\put(-19,+4){${\cal L}_{0}$}
\put(+56,+3){${\cal L}$}
\put(+82,+21){${\cal L}_{\infty}$}

\put(0,0){\oval(160, 50)}
\put(+20,+27){\vector(-1,0){10}}
\put(+20,+4){\vector(-1,0){10}}
\put(-29,+1){\vector(+1,+1){6}}

\end{picture}
\vspace{25mm}

\begin{center}
Fig. 1  (3-connected domain and Cauchy  theorem)
\end{center}

\noindent
Besides, all functions  $z^{-1}\; Q_{n}(z)$ are continuous  up to the  boundary ${\cal L}^{all}$, hense
Cauchy theorem will hold in this 3-connected domain:
$$
- \; \oint _{{\cal L}} \; Q_{n}(z) \;{dz \over z} \; +  \;
\oint _{{\cal L}(0)} \; Q_{n}(z) \;{dz \over z} \; + \;
\oint _{{\cal L}(\infty)} \; Q_{n}(z) \;{dz \over z} \; = 0 \; ;
\eqno(2.10a)
$$

\noindent
where the sign minus before the first integral  refers to  "improper" \hspace{2mm}
direction of going around the contour ${\cal L}$.
With the use of  conventional residue notation,  from (2.10a) we arrive at the main working
formula
$$
\oint _{\cal L} \; Q_{n}(z) \;{dz \over z} \; = -\; (2\pi i)
\sum_{z=0,\infty} \; res \;   { Q_{n}(z) \over  z}  \; .
\eqno(2.10b)
$$

\noindent
Thus, for the zero order term in (2.9b) we obtain
$$
\oint_{{\cal L}} \; Q_{0}(z) \; {dz \over z} =
- \; (2\pi i) \; \sum_{z=0,\infty} \; \; res \;
{\sqrt{2\epsilon M\;z^{2} + 2\alpha M\;z -L^{2}}\over z} =
$$
$$
= 2 \pi  \;  (\;  - L  + i\; {\alpha M \over \sqrt{2\epsilon M}}  \; ) \; .
\eqno(2.11)
$$

\noindent
As for  the next order term  $Q_{1}(t)$ we have
$$
Q_{1}(t) = - \; { 1 \over 2Q_{0}(t)} \; Q'_{0}(t) =
-{1 \over 2} \; {(2\epsilon M\; e^{2t} + \alpha M \; e^{t}) \over
 (2\epsilon M\; e^{2t} + 2\alpha M \; e^{t}  - L^{2}) } \; ,
\eqno(2.12a)
$$

\noindent and its contribution is
$$
{\hbar \over i} \; \oint_{{\cal L}} \; Q_{1}(z) \; {dz \over z} =
- \; \pi \; \hbar   \; .
\eqno(2.12b)
$$

\noindent
With regard to $Q_{0}$   and   $ Q_{1} $, eq.   (2.9b)  gives the following
($L = \hbar \; (l+{1\over 2})$ )
$$
- L  + i\; {\alpha M \over \sqrt{2\epsilon M}}  -
{ \hbar \over 2}  =  \hbar \; n \;
\qquad
\; \Longrightarrow \; \qquad
\epsilon = - {\alpha^{2} M \over 2 \hbar^{2} N ^{2}}   \; .
\eqno(2.13)
$$

\noindent         where $N = (n+l+1) $. The energy spectrum obtained  coincides with
the known exact result from the Schr\"{o}dinger  quantum mechanical equation.

Now it is the point to understand  this coincidence from a more general viewpoint.
What are the reasons for such a success of the seemingly approximate technics.
The form of $Q_{0}$ as a second order polynomial  is  relevant point; and another essential fact
is that  ${\cal L}$-contour encloses  zeros  (classical turning points or branch points of $\sqrt{P}$) of this polynomial.
At these  general assumptions we always will obtain
$$
Q_{0}(t) =  \sqrt{ P} \; , \qquad
P = ( A e^{2t} + B e^{t} +C ) \; ,
\eqno(2.14)
$$
$$
res \;_{\;\;z=0}   { Q_{0}(z)\over z}    = \sqrt{C} \; , \qquad
res \;_{\;\;z=\infty }  { Q_{0}(z)\over z}  =
-\; {B \over 2\sqrt{A}}  \;  \Longrightarrow
$$
$$
\oint _{{\cal L}} \;Q_{0}\; dt = 2\pi \;  [
-\sqrt{-C} \; + i \;   {B \over 2\sqrt{A}}  \; ] \; ;
$$
$$
Q_{1}(t) = -{Q_{0}' \over 2Q_{0}} =
-{1\over 2} \; {P'\over 2\sqrt{P}} \; {1\over \sqrt{P}} = -{1\over 2}\;
{P' \over 2P}   \; ,
\qquad  \Longrightarrow
$$
$$
{\hbar \over i}\; \oint _{{\cal L}} \;Q_{1}\; dt =  -{\hbar \over 2}\; 2\pi \; .
$$

\noindent
Now we are going to elucidate general structure of all terms of  greater order   than the first.
For  $Q_{2}(t)$ we have
$$
Q_{2}(t) =  -{1 \over 2 Q_{0}} \; ( \; Q_{1}' + Q_{1}^{2} \; ) =
$$
$$
=   - \; {1 \over 2 \sqrt{P}} \; [ (
\; (-{1\over 2} ) {P'' \over 2P } + ({1\over 2})\;{P' P' \over 2P^{2} } ) +
({1\over 4} )\; {P' P' \over 2^{2} P^{2}} \;  ] =
$$
$$
=    -\; {1 \over 2 \sqrt{P}} [   \;
-\; { P'' \over 2^{2} P} + {5\over 4} \; {P' P' \over 2^{2} P^{2}} \; ] \; .
\eqno(2.15)
$$

\noindent
Further it will be helpful to use special designation. The terms  $Q_{1} , \; Q_{2}$
let be  expressed symbolically as
$$
Q_{1} \sim {2' \over 2} \; , \qquad
Q_{2} \sim
{1 \over  \sqrt{P}}  (   \;
{ 2' \over 2}  \; \oplus   \; {4' \over 4 } \;  ) \; ,
\eqno(2.16a)
$$

\noindent where
$$
( -{P' \over 2P}  ) \;\; \mbox{refers to}\;  \;\;  как \;\; {2' \over 2}\;\; ,
$$
$$
( -\; { P'' \over 2^{2} P} ) \;\; \mbox{refers to }  \;\; { 2' \over 2}  \; \; ,
$$
$$
({1\over 4} )\; {P' P' \over 2^{2} P^{2}} \;\;\; \mbox{refers to} \; \; \;
{4' \over 4 }     \; .
$$

\noindent In other words, the  notation used   keeps only information relevant to
further calculating the residues $res_{0}$ and $res_{\infty}$:

\begin{quotation}

1) we indicate the power of polynomials at numerator and denominator of any term;

2) also we indicate  that the numerator  - polynomial in $e^{t}$ was differentiated
with respect to $t$-variable  and therefore the constant term of the
polynomial is suppressed.

No matter how many times the  polynomial was differentiated, the same  single symbol $'$ stands for all
these cases.

\end{quotation}

In such symbolic notation the  operator $d / dt$ acts as follows
$$
{d \over d t} \; {1 \over \sqrt{P}} = -\; {1 \over 2\sqrt{P}} \; {P' \over P}
\sim  {1 \over \sqrt{P}} \; {2' \over 2}  \;\; , \;\;\;
$$
$$
{d \over d t} \; {2' \over 2}\;   \sim \;
[\; {2' \over 2} \oplus {4' \over 4} \; ] \;\; ,
$$
$$
{d \over d t} \; {4' \over 4}\;   \sim \;
[\; {4' \over 4} \oplus {6' \over 6} \; ] \; , ...
$$

\noindent
Now we are ready to proceed to  $Q_{3}$ term:
$$
Q_{3} = -\; {1 \over 2 Q_{0} } \; ( Q_{2}' + 2 Q_{1} Q_{2} )  \sim
 \;
 $$
 $$
 \sim \;
{1 \over \sqrt{P}} \; \left  \{  [
{d \over d t}
{1 \over  \sqrt{P}} (   \;
{ 2' \over 2}  \; \oplus   \; {4' \over 4 } \;  )    ]
\; + \;  {2' \over 2}   \;
{1 \over  \sqrt{P}} (   \;
{ 2' \over 2}  \; \oplus   \; {4' \over 4 } \;  )  \right \} \sim
$$
$$
\sim \; {1 \over P} \left [
{2'\over 2} ( { 2' \over 2}  \; \oplus   \; {4' \over 4 }  ) \oplus
{d \over d t} ( { 2' \over 2}  \; \oplus   \; {4' \over 4 }  )
\oplus {2' \over 2}  ( { 2' \over 2}  \; \oplus   \; {4' \over 4 }
) \right ] \; .
$$

\noindent So, the final structure of  $Q_{3}$ looks as
$$
Q_{3} \sim \;
 {1 \over P} \;  (
{2' \over 2}  \; \oplus \; {4' \over 4}  \; \oplus \; {6' \over 6}  ) \; .
\eqno(2.16b)
$$

\noindent
Let us consider else one term:
$$
Q_{4} = - {1 \over 2Q_{0}} \; (Q_{3}' + 2Q_{1}Q_{3} +Q_{2}^{2} ) \; .
$$

\noindent With the use of symbolic notation we get to
$$
Q_{4}   \sim  \; {1 \over \sqrt{P}} \;  \{ \;
{d \over dt}
[\; {1 \over P} \;  (
{2' \over 2}  \; \oplus \; {4' \over 4}  \; \oplus \; {6' \over 6}  )
\; ]\; \oplus
\eqno(2.16c))
$$
$$
\oplus {2'\over 2} \;\; {1 \over P} \;  (
{2' \over 2}  \; \oplus \; {4' \over 4} \; \oplus \; {6' \over 6}  )
\; \; \oplus    \; \;
[ \; {1 \over  \sqrt{P}}  (   \;
{ 2' \over 2}  \; \oplus   \; {4' \over 4 } \;  ) \;  ]^{2}  \;  \} \sim
$$
$$
\sim {1 \over P\sqrt{P}} \;  [ \;
{ 2' \over 2}  \; \oplus   \; {4' \over 4 } \; \oplus \; {6' \over 6 }
\oplus \; {8' \over 8 }  \; ] \; .
$$

\noindent
At this point we are ready to write down symbolically the following general structure
of an arbitrary order term $Q_{n}$ of WKB-series:
$$
Q_{n} = {1 \over (\sqrt{P})^{n-1}} \;
\left [\;  { 2' \over 2}  \; \oplus   \; {4' \over 4 } \; \oplus \; ... \oplus \;
 {(2n)'\over (2n) } \; \right   ]   \; .
\eqno(2.17a)
$$

\noindent It is readily verified that all required residues vanish  identically:
$$
res \;_{\;\;z=0} \;   {Q_{n}(z) \over z }  =
{1 \over (\sqrt{C})^{n-1}} \; 0 = 0 \;\;  , \;
$$
$$
res\;_{\;\;z=\infty} \;   {Q_{n}(z) \over z}   =
- res\;_{\;\;y=0} \;   {Q_{n}(y^{-1}) \over y }    =
\eqno(2.17b)
$$
$$
= - \; res \;_{\;\;y=0} \; {1 \over y\; (\; \sqrt{A y^{-2}  +B y^{-1} + C } \; )^{n-1} }
\; [
{ 2' \over 2}  \; \oplus   \; {4' \over 4 } \; \oplus \; ... \oplus \;
{(2n)'\over (2n)}   ]_{z = y^{-1}} =
$$
$$
= -\; res \;_{\;\;y=0} \; {y^{n-1} \over y \; (\; \sqrt{A +B y^{-1} + C y^{-2} } \;)^{n-1} }
\;\; const \; = 0 \; .
\eqno(2.17c))
$$

\noindent
So, the success of Borh-Sommerfeld quantization for the  hydrogen atom is rationalized in full.

\subsection*{
3. Non-relativistic hydrogen atom in Lobachevsky space $H_{3}$}

\begin{quotation}

In Section 3 it is shown that the
Borh-Sommerfeld quantization as applied to the  hydrogen atom in
Lobachevsky space $H_{3}$ provides us with an exact energy spectrum of this generalized atom model.

\end{quotation}

In Lobachevsky space the radial wave equation has the form
$$
{1 \over \sinh^{2} }\;{d \over d \chi } \sinh^{2} \chi {d \over d \chi} \; R +
[ \; {2M\rho^{2} \over \hbar^{2}} \;  ( \epsilon +
{\alpha \over \rho \tanh \chi }  ) - {l(l+1) \over \sinh^{2} \chi } \;
] \; R (\chi ) = 0  \; .
\eqno(3.1)
$$

\noindent Introducing a new variable  $ t = \ln (\tanh \chi) $
$$
{d^{2} \over dt^{2}} R  +  {d \over dt }  R  +
{1 \over (1 - e^{2t})^{2}}
[ {2\epsilon  \rho^{2} M e^{2t} + 2 \alpha \rho^{2} M  e^{t} \over \hbar^{2} } -
l(l+1)(1 - e^{2t}) ] R = 0 \;
\eqno(3.2a)
$$

\noindent and excluding the first-derivative term by the substitution $S = e^{t/2}\;R$ one gets to
$$
{d^{2} \over dt^{2}} \; S \;  + \;
\left [ \; {1 \over (1 - e^{2t})^{2}}  \;
\left ( \; {2\epsilon  \rho^{2} M \; e^{2t} + 2 \alpha \rho^{2} M \; e^{t} \over \hbar^{2} } -
\; l(l+1)(1 - e^{2t})\; \right ) -{1 \over 4} \;\right  ] \; S = 0 \; .
\eqno(3.2b)
$$

\noindent
Further we are forced to move through  trial and error method.
Because we have known an exact energy spectrum of the system in  advance,
we are going to proceed in the following way:
in the exact  quantization condition  let us take into account only two first terms
of a certain formal WKB-series, on obtaining the required  exact result we will
analyze next terms.

Firstly, with the use of the trick
$$
-\; {l(l+1)  \over 1-e^{2t}} = {1\over  1-e^{2t}}
[\; -(l+{1 \over 2} )^{2} + {1\over4 } \; ]  =
{1\over  1-e^{2t}} \; ( - {L^{2} \over \hbar^{2}}  +{1 \over 4} ) \; ;
$$

\noindent eq.  (3.2b)  becomes
$$
{d^{2} \over dt^{2}}  S   +
\left [
 \; {2\epsilon  \rho^{2} M \; e^{2t} + 2 \alpha \rho^{2} M \; e^{t}
-L^{2}(1 - e^{2t})   \over \hbar^{2} (1 - e^{2t})^{2} }
\;  +  {1 \over 4}  ( {1 \over 1-e^{2t}} -1  ) \right ] \; S = 0 \; ,
\eqno(3.3a)
$$

\noindent or shorter
$$
{d^{2} \over dt^{2}}\; S(t) \; + \; [ \; {\Pi^{2}(t) \over \hbar^{2}} \;+
\; \Delta (t)  \; ] \;S (t) = 0 \; ,
\eqno(3.3b)
$$
$$
\Pi^{2}(t) = {(2\epsilon  \rho^{2} M + L^{2})  e^{2t} +
2 \alpha \rho^{2} M \; e^{t}  - L^{2} \over (1-e^{2t})^{2} } \; , \qquad
\Delta (t) = {1 \over 4} {e^{2t}\over (1-e^{2t})} \; .
$$

\noindent Further we act as in Section 2:
$$
S(t) = exp [ \; {i\over \hbar } \; \int \; Q(t) dt \;  ] \; ,\qquad
{\hbar \over i} \; {d \over dt} Q + Q^{2} - \Pi^{2} +
( {\hbar \over i}) ^{2} \Delta = 0\; , \;\;
$$
$$
Q(t) = \sum_{n=0}^{\infty} \;  [ \;
({\hbar \over i})^{n} \; Q_{n}(t)\; ] \; , \qquad
Q_{0} = \sqrt{\Pi^{2}} \; , \; Q_{1} = -{1 \over 2Q_{0}} \; Q'_{0} \; , \;\;
$$
$$
Q_{2} = -{1 \over 2Q_{0}}\; [ \;Q'_{1} + Q_{1}^{2} + \Delta \; ] \; , ...
$$
$$
Q_{n} = -\; {1 \over 2 Q_{0}} \;
[ \; {d \over dt} Q_{n-1} \; + \;
\sum_{k=1}^{n-1}  \; Q_{n-k}  Q_{k} \;  ] = 0 \; , \; n = 3,4,5, ...
\eqno(3.4a)
$$

\noindent
Quantum condition is the same
$$
{1 \over 2\pi i} \; \oint_{{\cal L}}
[{d \over d t} \ln S(t) ] \; dt = n \;  \; \Longrightarrow \;\;
$$
$$
\sum_{n=0}^{\infty} \; [ \; ({\hbar \over i})^{n} \;
\oint _{\cal L} \; Q_{n}(t) \;dt \;   ] \; = 2 \pi \hbar \; n \; .
\eqno(3.4b)
$$

\noindent In contrast to previous case (Section 2) here  we are to
employ  Cauchy theorem for a  5-connected domain:

\vspace{+5mm}
\unitlength=0.6mm
\begin{picture}(160,40)(-140,0)
\special{em:linewidth 0.4pt}
\linethickness{0.4pt}

\put(-120,0){\line(+1,0){128}}
\put(+52,0){\vector(+1,0){40}}

\put(+8,0){\line(+1,0){2}}
\put(+12,0){\line(+1,0){2}}
\put(+15,0){\line(+1,0){2}}
\put(+18,0){\line(+1,0){2}}
\put(+21,0){\line(+1,0){2}}
\put(+24,0){\line(+1,0){2}}
\put(+27,0){\line(+1,0){2}}
\put(+30,0){\line(+1,0){2}}
\put(+33,0){\line(+1,0){2}}
\put(+36,0){\line(+1,0){2}}
\put(+39,0){\line(+1,0){2}}
\put(+42,0){\line(+1,0){2}}
\put(+45,0){\line(+1,0){2}}
\put(+48,0){\line(+1,0){2}}
\put(+51,0){\line(+1,0){2}}
\put(+54,0){\line(+1,0){2}}

\put(+90,-5){$r$}
\put(-20,-40){\vector(0,+1){80}}

\put(+5,-2){\line(+1,0){50}}
\put(+5,-2){\line(0,+1){4}}
\put(+55,+2){\line(-1,0){50}}
\put(+55,-2){\line(0,+1){4}}

\put(+7,0){\circle*{2}}
\put(+53,0){\circle*{2}}
\put(-20,0){\circle*{2}}
\put(-20,0){\circle{7}}

\put(-110,0){\circle{7}}
\put(+70,0){\circle{7}}
\put(-110,0){\circle*{2}} \put(-110,+4){${\cal L}_{-1}$}
\put(+70,0){\circle*{2}}  \put(+70,+4){${\cal L}_{+1}$}

\put(-19,+4){${\cal L}_{0}$}
\put(+56,+3){${\cal L}$}
\put(+98,+18){${\cal L}_{\infty}$}

\put(-20,0){\oval(260, 50)}
\put(+20,+27){\vector(-1,0){10}}
\put(+20,+4){\vector(-1,0){10}}
\put(-29,+1){\vector(+1,+1){6}}

\end{picture}
\vspace{25mm}

\begin{center}
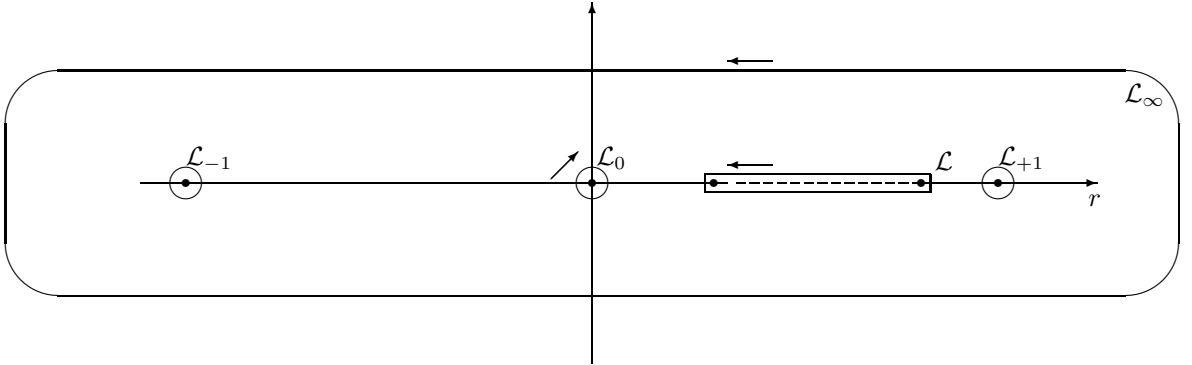
Fig. 2  (5-connected domain and Cauchy theorem)
\end{center}

Cauchy theorem provides us with relation
$$
- \; \oint _{{\cal L}} \; Q_{n}(z) \;{dz \over z} \; +  \;
\eqno(3.5a)
$$
$$
+ \;
\oint _{{\cal L}(0)} \; Q_{n}(z) \;{dz \over z} \; + \;
\oint _{{\cal L}(-1)} \; Q_{n}(z) \;{dz \over z} \; + \;
$$
$$
+ \;
\oint _{{\cal L}(+1)} \; Q_{n}(z) \;{dz \over z} \; + \;
\oint _{{\cal L}(\infty)} \; Q_{n}(z) \;{dz \over z} \; = 0 \; ;
$$

\noindent the contour ${\cal L}$ is going around in improper direction which reflects
the sing minus at the first term. From (3.5a) it follows
$$
\oint _{\cal L} \; { Q_{n}(z) \over z} \;dz \;  = ( - 2 \pi i )\;
\sum \;res\; _{\;\;z = 0, \pm 1, \infty} \; ( { Q_{n}(z) \over z} ) \; .
\eqno(3.5b)
$$

\noindent
Two first terms of the WKB-series contribute to quantum condition  (to do formulae shorter
dimensionless  quantities are used)
$$
\oint _{{\cal L}} \; Q_{0}(t) \; d t = (- 2 \pi i ) \;
\left [\;  \sqrt{-L^{2}}  \; - \; {1 \over 2} \; \left (\; \sqrt{
2 \epsilon  +  2 e } +  \sqrt{ 2 \epsilon - 2 e } \; \right ) \right ] \; , \;\;
$$
$$
{\hbar \over i } \; \oint_{{\cal L}} \; Q_{1}{dz \over z} =
 - \pi \; \hbar \; \;  .
\eqno(3.6)
$$

\noindent
which results in
$$
{i \over 2} \;  ( \sqrt{
2\epsilon   + 2 e  } +
\sqrt{ 2\epsilon   - 2 e }  ) =
\hbar \; (n +l+1)  \qquad  \Longrightarrow \;
$$
$$
N = n+l+1, \qquad \epsilon =  - \; {e^{2} \over 2N^{2}} - {N^{2} \over 2}\; .
\eqno(3.7a)
$$

\noindent However we have known an exact energy formula [47]
$$
\epsilon =  - \; {e^{2} \over 2N^{2}} - {N^{2} -1 \over 2}\;.
\eqno(3.7b)
$$

\noindent So, (3.7a), what we have found,   slightly differs from expected eq. (3.7b); difference
is in an additive term $ {1 \over 2}$.
Returning to the expression (3.3b) for  $\Pi^{2}$, one  notices that it would  suffice a single change
$
(2\epsilon  + L^{2}) \; \Longrightarrow \; $
$
(2\epsilon  - \hbar^{2}  + L^{2})     \; ,
$
to reproduce through 2-term WKB-quantization just the exact energy formula (3.7b).
This change would  lead to
$$
  \Delta = {e^{2t} \over 4(1 -e^{2t})} \qquad \Longrightarrow \qquad
( {e^{2t} \over 4(1 -e^{2t})}   + {e^{2t} \over (1 -e^{2t})^{2} } ) =
{5 - e^{2t} \over 4} \;  {e^{2t} \over (1 -e^{2t})^{2}} \; .
$$

\noindent
In other  words, instead formulas (3.3b) one should begin with the following equations
$$
{d^{2} \over dt^{2}}\; S(t) \; + \; [ \; {\Pi^{2}(t) \over \hbar^{2}} \;+
\; \Delta (t) \;   ] \;S (t) = 0 \; ,
\eqno(3.8a)
$$
$$
\Pi^{2}(t) = {(2\epsilon   - \hbar^{2} + L^{2})  e^{2t} +
2 \alpha \; e^{t}  - L^{2} \over (1-e^{2t})^{2} } \; ,
$$
$$
\Delta (t) =
{5 - e^{2t} \over 4} \;  {e^{2t} \over (1 -e^{2t})^{2}} \; .
\eqno(3.8b)
$$

For the following it will be convenient to have solved the task (3.8b) in a general form
$$
Q_{0}(t) = \sqrt{\Pi^{2}(t)} =
{\sqrt{A e^{2t} + B e^{t} + C } \over 1 - e^{2t} } =
{\sqrt{P(t)} \over 1 -e^{2t}} \; ,
$$

\noindent and further
$$
res \;_{\;\;z=0}\; {Q_{0}\over z } = \sqrt{C} \; , \qquad
res \;_{\;\;z= \infty }\; {Q_{0}\over z } = 0 \; ,
$$
$$
res \;_{\;\;z= +1 } \; {Q_{0}\over z } = -\;{1\over 2}\; \sqrt{A+B +C}\; ,
$$
$$
res \;_{\;\;z= -1 } \; {Q_{0}\over z } = -\;{1\over 2}\; \sqrt{A-B +C}\; , \;\;
$$
$$
\oint_{{\cal }} \; Q_{0}{dz \over z} =
2\pi  [\;  - \sqrt{-C} + {i\over 2} \;
( \sqrt{A+B +C} + \sqrt{A-B +C} ) \; ]  \; ,
\eqno(3.9a)
$$
$$
Q_{1} = -\; {1\over 2}  ( {P' \over 2 P } + {2e^{2t} \over 1 - e^{2t}} ) \; ,\qquad
{\hbar \over i}\;
\oint_{{\cal }} \; Q_{1}{dz \over z} =
- \; 2\pi \; {\hbar \over 2} \; .
\eqno(3.9b)
$$

\noindent
The 2-term WKB-quantization provides us with
$$
\left [\; - \sqrt{-C} +
{i\over 2} \; (\sqrt{A+B+C} + \sqrt{A-B+C}) \; \right ]\;  - {\hbar \over 2} =
\hbar \; n    \; ,
\eqno(3.9c))
$$

\noindent as applied to (3.8b) this leads to
$$
i\; {1 \over 2} \; \left ( \sqrt{
2\epsilon -1  + 2 e } +
\sqrt{ 2 \epsilon - 1  - 2e  } \right ) = N \; , \;\; N = (n +l+1) \; ,
$$

\noindent from which it follows an exact energy formula.
It can be readily verified by a straightforward calculation that the next term $Q_{2}$ (see
(3.4a) and (3.8b))  gives a zero to the main quantization condition\footnote{
General task of  proving  that all terms  greater order than $Q_{2}$ do not contribute to
the quantum condition looks quite interesting from mathematical viewpoint, but it will
 be omitted here.}

\subsection*{
4. Hydrogen atom in Riemann space of constant curvature $S_{3}$}

To have analyzed   the  hydrogen atom in the same line on the Riemann constant
curvature background we  need no new ideas in addition to previous case (Section 3).
And so it suffices to write down  only several main relations:

Radial wave equation has the form
$$
{1 \over \sin^{2} }\;{d \over d \chi } \sin^{2} \chi {d \over d \chi} \; R +
[ {2M\rho^{2} \over \hbar^{2}} \; ( \epsilon +
{\alpha \over \rho \tan \chi }  ) - {l(l+1) \over \sin^{2} \chi } \;
] \; R (\chi ) = 0  \; .
$$

\noindent Quantization condition remains   the same.
Computation of contour integrals is reduced to four  residues:
$$
\oint _{\cal L} \; { Q_{n}(z) \over z} \;dz \;  = ( - 2 \pi i )\;
\sum \;res\; _{z = 0, \pm i, \infty} \; ( { Q_{n}(z) \over z} ) \; .
$$

\noindent 2-term  quantization gives
$$
{i \over 2} \;  ( \sqrt{
-(2\epsilon \rho^{2} M + \hbar^{2})  + 2i \alpha \rho M  } +
\sqrt{ -( 2 \epsilon \rho^{2} M    - 2i \alpha \rho M } ) =
\hbar \; (n +l+1)  \; ,
$$

\noindent
from which it follows the exact energy formula
$$
\epsilon =  - \; {e^{2} \over 2N^{2}} + {N^{2} -1 \over 2}\;.
$$

\subsection *{
5. WKB-analysis of the  Klein-Fock equations in spaces   $E_{3},H_{3},S_{3}$ }

\begin{quotation}

The hydrogen atom models based on Klein-Fock equation in three spaces $E_{3},H_{3},S_{3}$
are described by radial equations which  come within the general types that have been considered at
non-relativistic case (see  (2.17) and (3.9)), hence we can act nearly formally without any serious
additional work. At this line one can readily produce three known exact energy spectra on the
 base of the 2-term quantization.

\end{quotation}

In flat space $E_{3}$
$$
\epsilon = mc^{2} \left ( 1 + {\alpha^{2}/\hbar^{2} c^{2} \over N^{2} } \right )^{-1/2}
\;\; .
\eqno(5.1)
$$

In the Riemann and Lobachevsky spaces (upper and lower sign correspondingly)
$$
\epsilon =  m c^{2} \left (  1 +  { \alpha^{2} / \hbar^{2} c^{2} \over N^{2} }
\right )^{-1/2}  \;
\sqrt{ 1 \pm {\hbar^{2} \over  m^{2} c^{2} \rho^{2} }
( N^{2} - 1 +  \alpha^{2} / \hbar^{2} c^{2} )  }  \;
\eqno(5.3)
$$

\noindent where
$
N =  (n_{r} +1/2  +
\sqrt{(l+1/2)^{2} - \alpha^{2} /  \hbar^{2}  c^{2} }\; $.

\subsection*{6. Energy spectrum of a free Dirac particle on the sphere $S_{3}$}

Omitting all details on separating the variables in the free  Dirac equation in space of
positive constant curvature $S_{3}$, let us start with a corresponding radial equation
$$
{d^{2} \over d \chi^{2}} f \; + \; [\; { \rho^{2} (\epsilon^{2} - m^{2} c^{4}) \over
c^{2} \hbar^{2} } - { k^{2} + k \cos \chi \over \sin^{2} \chi } \; ] \; f = 0
\eqno(6.1)
$$

\noindent With a new variable  $t = \ln \; \tan \chi $, eq. (6.1) takes on the form
$$
{d^{2} \over dt^{2}}\; f \; - \; {1 - e^{2t} \over 1 + e^{2t}} \; {d\over dt} \; f \; +
$$
$$
+
{1 \over \hbar^{2}} \; {
[ { \rho^{2} (\epsilon^{2} - m^{2}c^{4})  \over c^{2}}   - L^{2} ] e^{2t}
-L^{2}  \over  (1 + e^ {2t} ) ^{2}     } \; f -
{ k \over (1+e^{2t} )^{3/2} }  \; f = 0 \; .
\eqno(6.2)
$$

\noindent
Further let us act in accordance with a standard scheme:
$$
f = exp [ {i \over \hbar} \int Q(t) dt ] \; , \qquad  Q(t) = \sum_{k=0}^{k=\infty}
({\hbar \over i})^{k} Q_{k}(t)  \; ,
$$
$$
{1 \over 2\pi i } \; \oint _{\cal L} {V'\over V} dt = n \;  , \qquad
\sum_{k=0}^{k=\infty} ({\hbar \over i})^{k}  \oint_{\cal L}Q_{k}(t) dt = 2\pi \hbar n
$$

\noindent For two first terms in the  quasi-classical series we have expressions:
$$
Q_{0}(t) = {
\sqrt{ [{\rho^{2} \over c^{2}} ( \epsilon^{2} - m^{2}c^{4}) - L^{2} ] e^{2t} - L^{2} }
\over
1 + e^{2t}  }
$$
$$
Q_{1}(t) = -{1 \over 2}  \left ( {Q_{0}' \over Q_{0}} -
{1 - e^{2t} \over 1 + e^{2t} } \right )
= - {1 \over 2} \; \left [
 { [{\rho^{2} \over c^{2} } \; (\epsilon^{2} - m^{2}c^{4}) - L^{2} ]\;  e^{2t} \over
[ {\rho^{2} \over c^{2} } \; ( \epsilon^{2} - m^{2} c^{4} ) - L^{2} ] e^{2t}
- L^{2} } -1 \right ]     \; ,
$$

\noindent their contributions are
$$
\oint_{\cal L} Q_{0} dt = 2\pi [ - k \hbar +
\sqrt{ {\rho^{2} \over c^{2}}
(\epsilon^{2} - m^{2} c^{4} )} ] \; ,
$$
$$
\oint_{\cal L} Q_{1} dt =  2 \pi (- {\hbar \over 2})
\eqno(6.3)
$$

\noindent So,  2-term  quantization gives
$$
\sqrt{{\rho^{2}\over c^{2}}  (\epsilon^{2} - m^{2} c^{4} )} =
(n + k + {1 \over 2} ) \quad \Longrightarrow
$$
$$
\epsilon = \sqrt{ m^{2}c^{4} + {c^{2}\hbar^{2} \over \rho^{2}}
(n + k + {1 \over 2})^{2}} \;
\eqno(6.4)
$$

\noindent which  is an exact energy spectrum for a free Dirac particle on the sphere $S_{3}$.

\subsection*{7. Dirac equation in  spaces  $H_{3}$ and $S_{3}$}

Let us consider  quasi-classical description of the hydrogen atom
on the base of Dirac equation in $H_{3}$-space.
After separation of variables in hyper-spherical coordinates   the radial system is
$$
{d f \over d \chi} + {k \over \sinh \chi } f - {\rho \over \hbar c }
( \epsilon + mc^{2} + {\alpha \over \rho } \tanh^{-1} \chi ) g
= 0\;, \;
$$
$$
{dg \over d \chi } - {k \over \sinh \chi} g + {\rho  \over \hbar c }
( \epsilon - mc^{2} + {\alpha \over \rho}  \tanh^{-1} \chi ) f
= 0\; ,
\eqno(7.1)
$$

\noindent so the task is reduced to a second order
differential equation
$$
{d^{2} \over d \chi^{2} } \; f  \; + \;
{ (\alpha / \rho ) \over \sinh \chi} \; { 1 \over [ {\alpha \over \rho} \cosh \chi +
(\epsilon + mc^{2})  \sinh \chi ] }\; \; {d \over d \chi }\; f \; +
\eqno(7.2)
$$
$$
+ \;    { \rho^{2} \over \hbar^{2} c^{2}} \; \left [\; (\epsilon +
{\alpha \over \rho } \tanh ^{-1} \chi  )^{2} \;  - m^{2} c^{4}\;  \right ] \;  - \;
{ k^{2} + k \cosh \chi \over
\sinh^{2} \chi } \; f \;  +
$$
$$
\; + \; { k \over \sinh^{2} \chi } \; \;
{ (\alpha / \rho ) \over   {\alpha \over \rho} \cosh \chi + (\epsilon + mc^{2})  \sinh \chi  }
\; f = 0 \; .
$$

\noindent
With a new variable
$
\ln (\tanh (\chi/2) ) = t$
this equation takes on the form
$$
{d^{2} \over dt^{2} } f \; + \;
\left [ { 1 - e^{2t} \over e^{2t}  + 2 a e^{t}  + 1 } -
{1 + e^{2t} \over 1 - e^{2t} } \right ] \; {d \over d t } f\; +
$$
$$
+ \; {1\over \hbar^{2}} \; {1 \over (1 - e^{2t})^{2} } \;
\left [ (2 e^{t} \; {\epsilon \rho \over c} \; + \;
{\alpha \over c}\; (1 + e^{2t})) ^{2}  \;
 - \; m^{2} \rho^{2} c^{4} 4 e^{2t} \; - \; L^{2} \; (1-e^{2t})^{2}
\right ] \; -
$$
$$
- \; k \; {2e^{t} \over 1 -e^{2t} }\; \; {a (1+e^{2t}) + 2e^{t} \over
(1+e^{2t} ) + a 2 e^{t} } \; f = 0 \; ;
\eqno(7.3a)
$$

\noindent
where
$$
L^{2} = \hbar^{2} k^{2} \; , \qquad  a = {\epsilon + mc^{2} \over \alpha / \rho }
\eqno(7.3b)
$$

\noindent
or in a short form
$$
f'' + V\; f' + ( { \Pi^{2} \over \hbar^{2} } + \Delta  ) f = 0 \; .
\eqno(7.4a)
$$

\noindent Further acting upon old scheme:
$$
f = exp  [  { i  \over \hbar }  \int Q(t) dt ] \; ,
$$
$$
 {\hbar \over i} Q' + Q^{2}
+ { \hbar \over i } V \; Q - \Pi^{2} + ( {\hbar \over i } ) ^{2} \Delta = 0
\eqno(7.4b)
$$

\noindent
for terms of quasi-classical series we find recursive relations:
$$
Q(t) = \sum _{n=0}^{n= \infty} ({\hbar \over i})^{n} Q_{n}(t) \; : \;\;
$$
$$
Q_{0}^{2} = \Pi^{2} \; , \;
Q_{1} = - { 1 \over 2 Q_{0}} ( Q_{0}'  + V \; Q_{0}) \; , \;\; ...
\eqno(7.4c))
$$

\noindent
Quantum condition is
$$
{1 \over 2\pi i } \; \oint _{\cal L} {f'\over f} dt = n \;\;\;  \Longrightarrow \;\;\;
\sum_{k=0}^{k=\infty} ({\hbar \over i})^{k}  \oint_{\cal L}Q_{k}(t) dt = 2\pi \hbar \; N
\eqno(7.4d)
$$

Here we  need  to discuss some new peculiarities. The matter is that the term
$Q_{0}$ in  $z = \tanh {\chi \over 2} $ variable has the form
$$
Q_{0} (z)  =
 \Pi (z) = {1 \over 1 - z^{2} } \;
\sqrt{ [ 2z \; {\epsilon \rho \over c } + {\alpha \over c }(1+z^{2}) ]^{2} -
m^{2} \rho^{2} c^{2} \; 4 z^{2} - L^{2} (1-z^{2})^{2}} \; ,
\eqno(7.5)
$$

\noindent which means that  under square root stays 4-order polynomial.
To make use the conventional residue theory we are forced to take for integration only
such contours that enclose all four branch points of $Q_{0}(z)$-term.

Let us show that two zeros of the function  $Q_{0}(z)$
are allocated in  physical region and two zeros are allocated in non-physical one.
To understand this let us consider in some detail such a relation
$$
\tanh \chi = { 2 \tanh {\chi\over 2} \over  1 + \tanh^{2} {\chi \over 2}}
\qquad \Longrightarrow  \qquad  w^{2} + 1 - {2w \over \tanh \chi } = 0 \;, \qquad w = \tanh {\chi\over 2} ;
$$

\noindent which has two solutions:
$$
w_{1} = z =
{ \cosh \chi - 1 \over \sinh \chi } \; , \qquad
w_{2} = Z =
{ \cosh \chi + 1 \over \sinh \chi } \; .
$$

\noindent  Taking into account relations
$$
z =   1 \; -  2 \; {1 - e^{-\chi} \over
e^{+\chi} - e^{-\chi}  } \; < \; 1  \;, \;
z = \tanh {\chi \over 2  }\; , \;
$$
$$
Z =  1 \; +  2 \; {1 + e^{-\chi} \over
e^{+\chi} - e^{-\chi}  } \; >  \; 1  \; , \qquad
Z \neq \tanh {\chi \over 2  }\; ;
$$

\noindent  one concludes that only one solution,   $z$ can be related with a physical variable
 $\tanh { \chi  \over 2}, \chi \in (0, + \infty) $; whereas  the second solution
  $Z \in (+1,+\infty )$
cannot be related in any sense to physical variable $\chi$.
Besides, between roots  $z$ and $Z$ there exists simple connection:
$$
w_{1} \; w_{2} = z\; Z = 1\; .
\eqno(7.6)
$$

\noindent
Now let us compute what the term $Q_{0}$ gives into quantum condition.
(7.4d):
$$
\oint _{\cal L} Q_{0}(t) dt \; ,
$$

\noindent  integration should be  taken along the contour enclosing classical turning points
on positive $\chi$-axis as shown in Fig. 3

\begin{center}
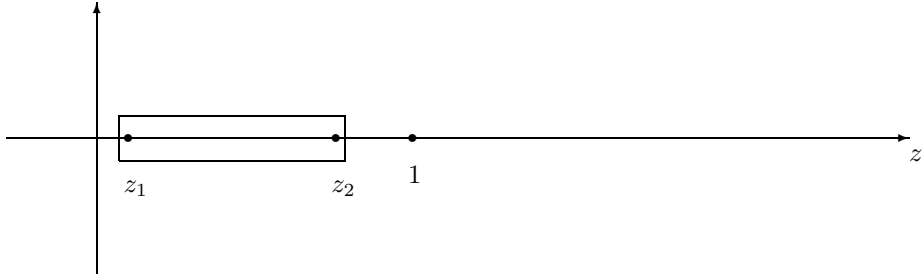
Fig 3. Physical integration contour  ${\cal L}(z)$.
\end{center}

\vspace{-7mm}
\unitlength=0.6mm
\begin{picture}(160,40)(-30,0)
\special{em:linewidth 0.4pt}
\linethickness{0.4pt}

\put(-20,0){\vector(+1,0){200}} \put(+180,-5){$z$}
\put(0,-30){\vector(0,+1){60}}

\put(+5,-5){\line(+1,0){50}}
\put(+5,+5){\line(+1,0){50}}
\put(+5,-5){\line(0,+1){10}}
\put(+55,+5){\line(0,-1){10}}

\put(+7,0){\circle*{2}}      \put(+6,-12){$z_{1}$}
\put(+53,0){\circle*{2}}     \put(+52,-12){$z_{2}$}

\put(+70,0){\circle*{2}}
\put(+69,-10){$1$}

\end{picture}
\vspace{20mm}

\noindent
However on mathematical reasons  it will be convenient  to change this integration contour
${\cal L} $ into doubled one ${\cal L'} $ (enclosing non-physical region of $\chi$-variable)

$$
\oint _{\cal L} Q_{0}(t) dt  =
{1 \over 2 } \;
\oint _{\cal L'} Q_{0}(t) dt  \; ;
\eqno(7.6c))
$$

\begin{center}
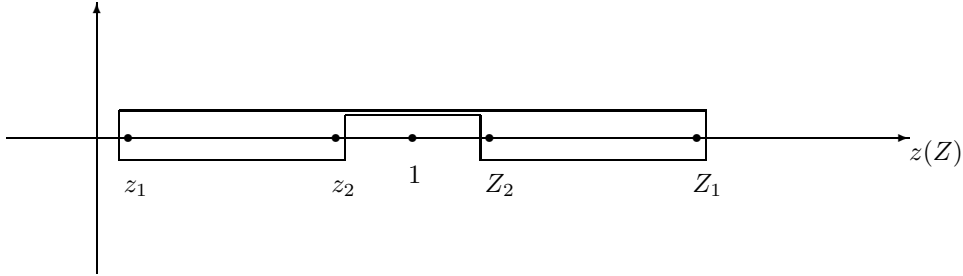
Fig. 4. Doubled integration contour  ${\cal L'}(z)$
\end{center}

\vspace{-7mm}
\unitlength=0.6mm
\begin{picture}(160,40)(-30,0)
\special{em:linewidth 0.4pt}
\linethickness{0.4pt}

\put(-20,0){\vector(+1,0){200}} \put(+180,-5){$z(Z)$}
\put(0,-30){\vector(0,+1){60}}

\put(+5,-5){\line(+1,0){50}}
\put(+5,-5){\line(0,+1){11}}

\put(+5,+6){\line(+1,0){130}}
\put(+135,+6){\line(0,-1){11}}

\put(+135,-5){\line(-1,0){50}}
\put(+85,-5){\line(0,+1){10}}
\put(+85,+5){\line(-1,0){30}}
\put(+55,+5){\line(0,-1){10}}

\put(+7,0){\circle*{2}}      \put(+6,-12){$z_{1}$}
\put(+53,0){\circle*{2}}     \put(+52,-12){$z_{2}$}

\put(+87,0){\circle*{2}}     \put(+86,-12){$Z_{2}$}
\put(+133,0){\circle*{2}}    \put(+132,-12){$Z_{1}$}

\put(+70,0){\circle*{2}}
\put(+69,-10){$1$}

\end{picture}
\vspace{20mm}

\noindent With such a trick we will count up total number of zeros: physical and unphysical ones.

\noindent  In so doing we will find contribution of the term $Q_{0}$ through conventional
 residue-technics:
$$
{1 \over 2 } \; \oint _{\cal L'} Q_{0}(t)  dt =
({1 \over 2}) \; (-2\pi i) \; \sum_{0,\infty,\pm 1} \; res \;
 {Q_{0} (z) \over z } \; .
\eqno(7.7a)
$$

\noindent
Allowing for
$$
res_{\;\;z=0} \; {Q_{0}(z) \over z} = Q_{0}(z=0) = \sqrt{{\alpha^{2} \over c^{2}}
- L^{2} }
$$
$$
res_{\;\;z=+1} \; {Q_{0}(z) \over z} =
-\sqrt{({\epsilon \rho \over c} + {\alpha \over c})^{2} - m^{2}\rho^{2}c^{2}} \; ,\;
$$
$$
res_{\;\;z=-1} \; {Q_{0}(z) \over z} =
+\sqrt{(-{\epsilon \rho \over c} + {\alpha \over c})^{2} - m^{2}\rho^{2}c^{2}}
$$
$$
res_{\;\;z= \infty} {Q_{0}(z) \over z} =
-
res_{\;\;y= 0} {Q_{0}(y^{-1}) \over y} =
 + \sqrt{ { \alpha^{2} \over c^{2}} - L^{2}}
$$

\noindent
1-term quantization provides us with
$$
({1 \over 2})\; (-2 \pi i ) \left [
2 \sqrt{{\alpha^{2} \over c^{2}} - L^{2}} +
\sqrt{
({\epsilon^{2} \rho^{2} \over c^{2}} - m^{2}\rho^{2}c^{2} +
{\alpha^{2} \over c^{2}}) + 2 {\alpha \epsilon \rho \over c^{2}} }
\;\;-     \right.
$$
$$
\left. - \; \sqrt{ ( {\epsilon^{2} \rho^{2}  \over c^{2}} - m^{2}\rho^{2} c^{2} +
{\alpha^{2} \over c^{2} }) - 2 {\alpha \epsilon \rho \over c^{2} } } \right ]
= 2\pi \hbar \; n
\eqno(7.7b)
$$

As for the next term contribution
$$
Q_{1} = -{1 \over 2Q_{0} } \;  [ \; {d \over dt} \; Q_{0}(t)  + V(t) \; Q_{0}(t)\;  ] \; .
\eqno(7.8a)
$$

\noindent with (in  units  $c=1, \hbar =1$)
$$
Q'_{0}(t) =
{+ 2 e^{2t} \over (1 - e^{2t})^{2} }
\; \sqrt{...} \; +
+ \; {1 \over (1 - e^{2t}) } \;
{1 \over  \sqrt{...}  }   \times
$$
$$
\times
\left [ [2\epsilon e^{t} + \alpha (1+e^{2t})]\; (2\epsilon e^{t} + \alpha 2 e^{2t}) -
m^{2} 4 e^{2t} \; + \; 2 L^{2}(1- e^{2t})  e^{2t} \right ] \; ;
\eqno(7.8b)
$$

\noindent for $(Q_{0}'/ Q_{0})$ anf  $V(t)$  one has
$$
{Q_{0}' \over Q_{0} } =
{2 e^{2t} \over 1 - e^{2t} }  +
  {
[2\epsilon e^{t} + \alpha (1+e^{2t})]\; (2\epsilon e^{t} + \alpha 2 e^{2t}) \;
 - \; m^{2} 4 e^{2t} \; +  \; 2 L^{2}(1- e^{2t}) e^{2t} \over
[2 \epsilon e^{t} + \alpha (1+ e^ {2t} ) ] ^{2}  \; -
\; m^{2} 4 e^{2t} \; - \; L^{2} (1-e^{2t})^{2}    }  \; ,
\eqno(7.8c)
$$
$$
V = {1 -e^{2t} \over e^{2t} + 2ae^{t} + 1 } -
{1 + e^{2t} \over 1 - e^{2t}} \; .
\eqno(6.8d)
$$

\noindent
Allowing for
$$
\oint _{\cal{L}}  \; Q_{1} (t) \; dt  =
-2 \pi i                   \; \times
$$
$$
\times   ( - {1 \over 2})
\left \{
\sum _{z = 0, \infty, \pm 1} \;  \; res \;
            \; \left ( {1 \over z}
{Q'_{0}(z) \over Q_{0}(z) }\; \right )  \; +  \;
\sum _{z = 0, \infty, \pm 1 }  \;  \; res \;
         \; \left (  {1 \over z} V (z) \right )  \right \} \; .
\eqno(7.9)
$$

\noindent
we arrive at the vanishing contribution of the term $Q_{1}$. Therefore,
2-term quantization leads to
$$
\epsilon = mc^{2}\;
\left ( 1 + {\alpha^{2} /\hbar^{2}c^{2} \over N^{2} } \right )^{-1/2}\;
\sqrt { 1 -
{\hbar^{2} \over m^{2} \rho^{2} c^{2} }   \;
( N^{2} +  \alpha^{2} / \hbar^{2} c^{2}) } \; .
\eqno(7.10)
$$

\noindent where
$
N =
n +  \sqrt{
k^{2} - \alpha^{2} / \hbar^{2} c^{2}} \; .
$

Consideration of  the  hydrogen atom on the base of Dirac equation in the space $S_{3}$
can be done in the same line. The final result is as follows
$$
\epsilon = mc^{2}\;
\left ( 1 + {\alpha^{2} /\hbar^{2}c^{2} \over N^{2} } \right )^{-1/2}\;
\sqrt { 1 +
{\hbar^{2} \over m^{2} \rho^{2} c^{2} }   \;
( N^{2} +  \alpha^{2} / \hbar^{2} c^{2}) } \; .
\eqno(7.11)
$$

\noindent where
$
N =
n +  \sqrt{
k^{2} - \alpha^{2} / \hbar^{2} c^{2}} \; .
$

\begin{quotation}

However, it must be stressed that these two formulas obtained are hardly correct in full
because the second one (7.11) does not result in an exact result at $\alpha =0$ (see Section 6)
\footnote{
This formula was found in [68] on the base of other considerations,
where it was considered as exact one.}. Moreover, turning to the next terms $Q_{2},Q_{3}$ and so on ,
we  might see that they give not-vanishing contributions. The latter indirectly proves only approximate
character of the formula (7.1). In should be added that till now  any  exact wave function solutions
of the latter task have not found.

\end{quotation}

\newpage

\subsection*{References}

\vspace{5mm}
\noindent
1.
Planck M.
{\em Die Quantenhypothese f\"{u}r  Molekeln mit  mehreren
Freiheitsgraden.}
Verhand\-lungen  der Deutschen  Physikalischen  Gesellschaft.
1915.  Bd. 17.  S. 407-418; S. 438-451.

\vspace{5mm}
\noindent
2.
Planck M.
{\em Die  physikalische  Struktur  des Phasenraum.}
Annalen der Physik.  1916. Bd. 50.   S. 385-418.

\vspace{5mm}
\noindent
3.
Wilson W. {\em  The quantum theory of  radiation and line spectra.}
Phil.  Magazine.  1915.  Vol. 29.  P. 795-802.

\vspace{5mm}
\noindent
4.
Ishiwara J. {\em Die universelle Bedeutung des Wirkungsquantums. }
Tokio Sugaku Buturigak\-kawi Kizi.  1915.  Bd 8.  S. 106-116.

\vspace{5mm}
\noindent
5. Sommerfeld A.  {\em Zur Theorie der
Balmerschen Serie.}  M\"{u}nchener  Berichte.  1915.  S. 425-458.

\vspace{5mm}
\noindent
6.
Sommerfeld A.
{\em Die Feinstructur  der wasserstoff und wasserstoff\"{a}hnlichen Linien.}
 M\"{u}nchener  Berichte.  1915.  S. 459-500;
{\em Zur Quantentheorie der Spektrallinien.}
 Annalen der Physik.  1916.  Bd 51.  S. 1-94; S. 125-167.

\vspace{5mm}
\noindent
7.
Schwarzschild K.  {\em Zur Quantenhypothese.}
 Berliner Berichte.  1916.  S. 548-568.

\vspace{5mm}
\noindent
8.
Epstein P.S.
{\it Zur Theorie des Starkeffektes.}
Z. Phys.  1916.  Bd. 17.  S. 148-150.

\vspace{5mm}
\noindent
9.
Epstein P.S.
{\em Zur Theorie des Starkeffektes.}
 Annalen der Physik.  1916.  Bd. 50.  S.489-520.

\vspace{5mm}
\noindent
10.
Epstein P.S.
{\em Zur Quantentheorie.}
 Annalen der Physik.  1916.  Bd. 51.  S.168-188.

\vspace{5mm}
\noindent
11.
Sommerfeld A.
{\em Atombau und Spektrallinien.}
Braunchweig, Vieweg, 1919.  S. 327-357; S. 520-522;

\vspace{5mm}
\noindent
12.
Wilson W.  {\em The quantum theory and electromagnetic phenomena.}
 Proc. Roy. Soc. London. A.  1922.  Vol. 102.  P. 478-483.

\vspace{5mm}
\noindent
13.
Wentzel G.
{\em Eine Verallgemeinerung der Quantembedingungen f\"{u}r die Zwecke der
Wellenmechanik.}
 Z. Phys.   1926.  Bd. 38.  S. 518-529.

\vspace{5mm}
\noindent
14.
Kramers H.A.    Z. Phys.   1926.  Bd. 39.  S. 828.

\vspace{5mm}
\noindent
15.
Brillouin M.L.
{\em La m\'{e}chanique ondulatoire  de Schr\"{o}dinger; une  m\'{e}thode
g\'{e}n\'{e}rale de r\'{e}solution par approximations successives.}
 Compt. Rend. Acad. Sci. Paris.  1926  Tom 183.  P. 24-44.

\vspace{5mm}
\noindent
16.
Brillouin M.L.
{\em Remarques sur la M\'{e}chanique ondulatoire.}
J. Phys.  Radium.  1926.   Vol. 7.  P. 353-368.

\vspace{5mm}
\noindent
17.
Dunham J.L.
{\em The Wentzel-Brillouin-Kramers metod of solving  the wave equation.}
 Phys. Rev.  1932.  Vol. 41.  P. 713-720.

\vspace{5mm}
\noindent
18.
Langer R.E.  Bull. Amer. Math. Soc.  1934.  Vol. 40.   P. 545.

\vspace{5mm}
\noindent
19.
Langer R.E.
{\em On the connection formulas and the solutions of the wave equation.}
 Phys. Rev. 1937.  Vol. 51.  P. 669-676.

\vspace{5mm}
\noindent
20.
Titchmarsh E.C.
Quart. J. Math.  1954.  Vol. 5.  P. 228.

\vspace{5mm}
\noindent
21.
Leller  J.B.
{\it Corrected Borh-Sommerfeld quantum conditions for nonseparable systems.}
 Ann. Phys. (N.Y.) 1958, Vol. 4, P. 180-188.

\vspace{5mm}
\noindent
22.
Bailey Paul B.
{\em Exact quantization rules for the one-dimensional Scr\"odinger
equation with turning points.}
 J. Math. Phys.  1964.  Vol. 5, No  9.  P. 1293-1297.

\vspace{5mm}
\noindent
23.
Froman N., Froman P.O.   {\em JWKB Approximation: contributions to the theory. }
North-Holland Publ. Co., Amsterdam, 1965.

\vspace{5mm}
\noindent
24.
Joseph B. Krieger. {\em Asymptotic properties of perturbation theory.}
J. Math. Phys.  1966.  Vol. 9, No  3.  P. 432-435.

\vspace{5mm}
\noindent
25.
Ponomarev L.I. {\em Lectures in quasi-classics} (in Russian). Kiev.  1966.

\vspace{5mm}
\noindent
26.
Krieger J.B. Rosenzweig C.
 Phys. Rev.  1967.  Vol. 164.   P. 171.

\vspace{5mm}
\noindent
27.
Krieger J.B., Lewis M.L.,  Rosenzweig C.
J. Chem. Phys.  1967.  Vol. 47.   P. 2942.

\vspace{5mm}
\noindent
28.
Krieger J.B.
J. Math. Phys.  1968.  Vol. 9.  P. 432.

\vspace{5mm}
\noindent
29.
Rosenzweig C., Krieger J.B.
{\em Exact quantization conditions.}
J. Math. Phys.  1968.  Vol. 9, No  6.  P. 849-860.

\vspace{5mm}
\noindent
30.
Voros A.
{\em Semi-classical approximations.}
Ann. Inst. H. Poincar\'{e}.  1976.  Vol.  A24.  P.31-90.

\vspace{5mm}
\noindent
31.
De Witt-Morette C.
{\em The semiclassical expansion.}
Ann. Phys.(N.Y.)  1976.  Vol. 97. P. 367-399.

\vspace{5mm}
\noindent
32.
Neveu A.
{\em Semiclassical quantization  methods in field theory.}
Phys. Rep. C.  1976.  Vol. 23. P. 265-272.

\vspace{5mm}
\noindent
33.
Mizrahi M.M.
{\em On the semiclassical expansion in quantum mechanics for arbitrary
Hamiltonians.}  J. Math. Phys.  1977.  Vol. 18.  P. 786-790.

\vspace{5mm}
\noindent
34.
Karl M. Bender, Kaare Olaussen and Paul S. Wang
{\em Numerological analysis of the WKB approximation in large order.}
Phys. Rev. D.  1977.  Vol. 16, No  6.  P. 1740-1748.

\vspace{5mm}
\noindent
35.
Gallas J.A.C.  Chem. Phys. Lett.  1983.  Vol. 96, No  1.  P. 479-480.

\vspace{5mm}
\noindent
36.
Comtet A., Bandrauk A.D., Campbell D.K.
{\em Exactness  of semiclassical  bound state energies for supersymmetric
quantum mechanics.}  Phys. Lett. B.  1985.  Vol. 150.  P. 159-162.

\vspace{5mm}
\noindent
37.
Dutt R., Khare A., Sukhatme U.P.
{\em Exactness of supersymmetric WKB spectra  for  shape-invariante
potentials. }  Phys. Lett. B.  1986.  Vol. 181.  P. 295-298.

\vspace{5mm}
\noindent
38.
Gomes M.A.F., Thomaz M.T., Vasconcelos G.L.
{\em Matrix formulation for the Wentzel-Kramers-Brillouin quantization rule.}
Phys. Rev. A.  1986.  Vol . 34, No  5.  P. 3598-3604.

\vspace{5mm}
\noindent
39.
Sch\"{o}pf  Hans-Georg.
{\em Zur Geschichte der Bohr-Sommerfeldschen Quantentheorie.}
 Ann. Phys.  1988. Bd. 45, No  8.  S. 595-604.

\vspace{5mm}
\noindent
40.
Kobylinsky N.A., Stepanov S.S., Tutik R.S.
{\em Semiclassical approach to ground state within the Klein-Gordon equation.}
J. Phys. A.  1990.  Vol. 23, No 6. P. 237-241.

\vspace{5mm}
\noindent
41.
Barclay D.T.,  Khare A. Sukhatme  U.
{\em Is the lowest order supersymmetric WKB
approximation exact for all shape invariant potentials?}
-- 8 pages.  hep-th/9311011.

\vspace{5mm}
\noindent
42.
Funahashi K., Kashiwa T.,  Sakoda S., Fujii K.
{\em Exactness in the Wentzel-Kramers-Brillouin  approximation for some
homogeneous spaces.}   J. Math. Phys.  1995.  Vol. 36.  P. 4590-4611;
 hep-th/9501145.

\vspace{5mm}
\noindent
43.
De Vega H.J., Larsen A.L., Sanchez N.
{\em Semi-classical quantization of circular strings in de Sitter and anti
de Sitter spacetimes.}
Phys. Rev. D.  1995.  Vol. 51.   P. 6917-6928;  hep-th/9410219.

\vspace{5mm}
\noindent
44.
Robnik M., Salasnich  L.
{\em WKB to all orders and the accuracy of the semiclassical quantization.}
J. Phys. A.  1997.  Vol. 30.  P. 1711-1718; quant-ph/9610027.

\vspace{5mm}
\noindent
45.
Robnik M., Salasnich  L.
{\em WKB exactness for the angular momentum and the Kepler
problem: from the torus quantization to the exact one.}
J. Phys. A.  1997.  Vol. 30.  P. 1719-1729; quant-ph/9603014.

\vspace{5mm}
\noindent
46.
Schr\"{o}dinger E.
{\em A method of determining quantum-mechanical eigenvalues and eigen\-functions.}
Proc. Roy. Irish. Soc. A.  1940.  Vol. 46, No  1.   P. 9-16.

\vspace{5mm}
\noindent
47.
Infeld L., Schild A.
{\em A note on the Kepler problem in a space of constant negative curvature.}
 Phys. Rev.  1945.  Vol. 67, No  2.  P. 121-122.

\vspace{5mm}
\noindent
48.
Bander M., Itzykson C.
{\em Group theory and the  hydrogen atom. I,II}
Rev. Mod. Phys.  1966.  Vol. 38.  P. 330-345; P. 346-357.

\vspace{5mm}
\noindent
49.
Aronson E.B., Malkin I.A., Manko V.I. {\em Dynamic symmetries in quantum theory.}
(in Russian). Elementary Particles and Atomic Nuclei. 1974. Vol. 5, No 1. P. 122-171.

\vspace{5mm}
\noindent
50.
Higgs P.W.
{\em Dynamical symmetries  in a spherical geometry. I.}
J. Phys. A.   1979.  Vol. 12, No  3. P. 309-323.

\vspace{5mm}
\noindent
51.
Leemon H.I.
{\em Dynamical symmetries  in a spherical geometry. II.}
 J. Phys. A.  1979.  Vol. 12, No  14.  P. 489-501.

\vspace{5mm}
\noindent
52.
Ringwood G.A. Devreese J.T.
{\em The hydrogen atom: Quantum mechanics on the quotient of a conformally
flat manifold.}  J. Math. Phys.  1980.  Vol 21.  P. 1390-1392.

\vspace{5mm}
\noindent
53.
Kurochkin Yu.A., Otchik V.S,
{\em Analogue of the Gunge-Lenz vector and energy spectrum in Kepler problem on
3-sphere.}  Doklady AN BSSR.  1979.   Vol. 23, No  11. P. 987-990.

\vspace{5mm}
\noindent
54.
Bogush A.A., Kurochlin Yu.A., Otchik V.S.
{\em On quantum-mechanical Kepler problem in Lobachevsky space.}
Doklady AN BSSR.  1980.  Vol. 24, No 1. P. 19-22.

\vspace{5mm}
\noindent
55.
Bessis N., Bessis G., Shamseddine R.
{\em Atomic fine structure in a space of constant curvature.}
J.  Phys. A.  1982.  Vol. 15, No 10.  P. 3131-3144.

\vspace{5mm}
\noindent
56.
Xu Chondming , Xu Dianyan.
{\em Dirac equation and energy levels of hydrogen-like
atoms in Robertson-Walker metric.}
Nuovo Cim. B. -- 1984.  Vol. 83, No 2. P. 162-172.

\vspace{5mm}
\noindent
57.
Bessis N., Bessis G., Shamseddine R.
{\em Space-curvature effects in atomic fine- and hyperfine-structure  calculations.}
Phys. Rev. A.  1984. Vol. 29, No 5. P. 2375-2383.

\vspace{5mm}
\noindent
58.
Duval C., K\"unzle H.
{\em Minimal gravitational coupling in the Newtonian theory and the covariant
Schr\"odinger equation.}
Gen. Rel. Grav. 1984.  Vol. 16. P. 333-347.

\vspace{5mm}
\noindent
59.
Shamseddine R.,
{\em Structure fine et hyperfine atomique dans un espace \`a courbure
constante. }  J. Phys. A. 1986.  Vol. 19, No 5.  P. 717-724.

\vspace{5mm}
\noindent
60.
Bessis N., Bessis G., Roux D.
{\em Space-curvature effects in the interaction between atom and external
fields: Zeeman and Stark effects in a space of constant positive curvature.}
Phys. Rev. A.  1988. Vol. 33, No 1.   P. 324-336.

\vspace{5mm}
\noindent
61.
Tagirov E.A.
{\em Quantum mechanics in Riemannian space-time. General covariant Schr\"odin\-ger
equation with relativistic corrections.}
 Dubna, 1988.  Preprint / E2-88-678.

\vspace{5mm}
\noindent
62.
Gorbatsievich  A.K., Priebe A.
{\em On the hydrogen atom in Kerr space time.}
Acta Phys. Polon. B.  1989.  Vol. 20, No 11. P. 901-909.

\vspace{5mm}
\noindent
63.
Groshe C. {\em The path integral for the Kepler problem on
the pseudosphere.}  Ann. Phys. (N.Y.)  1990.  Vol. 204. P. 208-222.

\vspace{5mm}
\noindent
64.
Katayama N.
{\em A note on the Kepler problem in a space of constant curvature.}
Nuovo Cim. B.  1990.  Vol. 105, Ser. 2, No 1.  P. 113-119.

\vspace{5mm}
\noindent
65.
Foerster A., Girotti H.O., Kuhn P.S.
{\em Nonrelativistic quantum particle in a  curved space as a constrained
system.} --   hep-th/9411052.

\vspace{5mm}
\noindent
66.
Gorbatcevich A.K., Tomilchik L.M. {\em Equation of motion of particles in conformally-flat space and
quark confinement.}
Minsk, 1986.  9 pages.  Preprint  No 415,  Belarus Academy of Siences, Institute of physics.

\vspace{5mm}
\noindent
67.
Melnikov V.N. Shikin G.N. {\em Hydrogen-like atom in gravitational field of the universe.}
 Izvestiya Vuzov. Fizika. 1985.  No 1.  P. 55-59.

\vspace{5mm}
\noindent
68.
Shamseddine R.
{\em On the resolution  of the wave equations of electron in a space of constant
curvature.}  Can. J. Phys.  1997.  Vol. 75, No  11. P.  805-811.

\end{document}